\documentclass{article}
\usepackage{spconf,amsmath,graphicx}
\usepackage{amsfonts}
\usepackage{booktabs, multirow}
\usepackage{tabularx}
\usepackage{xurl}
\usepackage{cite}
\usepackage{threeparttable}

\title{ECHO: Frequency-aware Hierarchical Encoding for Variable-length Signals}
\name{
  Yucong Zhang$^{1,3}$, 
  Juan Liu$^{2,1\dag}$, 
  Ming Li$^{2,3\dag}$\thanks{$^{\dag}$Corresponding authors: Juan Liu and Ming Li.}
}
\address{
  $^1$School of Computer Science, Wuhan University, Wuhan, China\\
  $^2$School of Artificial Intelligence, Wuhan University, Wuhan, China\\
  $^3$Suzhou Municipal Key Laboratory of  Multimodal Intelligent Systems, \\Digital Innovation Research Center, Duke Kunshan University, Suzhou, China\\
  \{yucong.zhang,liujuan\}@whu.edu.cn, ming.li369@dukekunshan.edu.cn
}

\begin{document}
%
\maketitle
\begin{abstract}
Pre-trained foundation models have demonstrated remarkable success in audio, vision and language, yet their potential for general machine signal modeling with arbitrary sampling rates—covering acoustic, vibration, and other industrial sensor data—remains under-explored.
In this work, we propose a novel foundation model ECHO that integrates an advanced band-split architecture with frequency positional embeddings, enabling spectral localization across arbitrary sampling configurations. Moreover, the model incorporates sliding patches to support inputs of variable length without padding or cropping, producing a concise embedding that retains both temporal and spectral fidelity and naturally extends to streaming scenarios.
We evaluate our method on various kinds of machine signal datasets, including previous DCASE task 2 challenges (2020–2025), and widely-used industrial signal corpora. Experimental results demonstrate consistent state-of-the-art performance in machine signal anomaly detection and fault classification, confirming the effectiveness and generalization capability of the proposed model. We open-sourced ECHO on \url{https://github.com/yucongzh/ECHO}.
\end{abstract}
\vspace{-2mm}\section{Introduction}\vspace{-2mm}
\label{sec:intro}
The reliable monitoring of machine health is critical for ensuring safety, reducing downtime, and optimizing operational efficiency across industrial domains. In recent years, machine signal analysis—encompassing acoustic emissions, vibration measurements, and other sensor modalities—has emerged as a central tool for detecting anomalous conditions and diagnosing faults before catastrophic failure. Traditional approaches, such as handcrafted feature extraction combined with conventional classifiers, have proven effective in narrow, domain-specific scenarios~\cite{yan2024comprehensive}. However, these methods often lack generalization capability across heterogeneous machine types, operating conditions, and sensing modalities.

Large-scale audio foundation models have emerged as a unifying paradigm for representation learning across diverse acoustic domains. Leveraging supervised~\cite{panns} or self-supervised~\cite{audioMAE, koutini22_interspeech, BEATs, CED, EAT} Vision Transformer~(ViT)-based~\cite{ViT} pre-training on large corpora, these models demonstrate strong transferability to downstream tasks ranging from tagging to captioning. Scaling training data further improves robustness across speech and audio tasks~\cite{Dasheng}. This success extends to industrial monitoring, where such models serve as front-end encoders for anomalous sound detection~(ASD)~\cite{jiang24c_interspeech, zheng2024improving, han2025exploring} and play key roles in recent DCASE Task~2 challenges. In domains with scarce labeled anomalies and varying conditions, the generalizable representations of audio foundation models provide a solid basis for domain adaptation.


\begin{figure*}[t]
  \includegraphics[width=\textwidth]{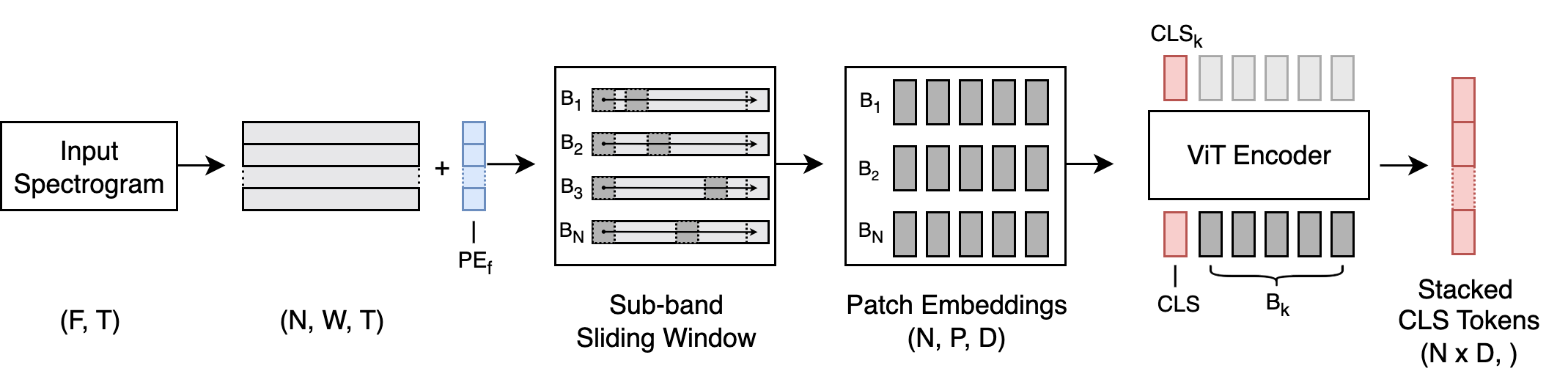}\vspace{-5mm}
  \caption{Feature extraction pipeline of the ECHO framework. F: number of frequency bins after STFT; T: number of time frames after STFT; N: number of sub-bands after band splitting; W: band width of sub-bands; P: number of sliding patches; D: feature dimension for each patch.}
  \label{fig:framework}
  \vspace{-5.5mm}
\end{figure*}

Despite their strengths, existing foundation models face two major challenges in real-world machine signal monitoring tasks. 
First, current ViT-based pre-trained models rely on fixed-size spectrogram input for patching, and adopt conventional 2D positional embeddings from image processing to learn 2D spatial relation among those patches. Modeling variable-length spectrogram thus requires truncation or interpolation, breaking the spatial relation between patches. We argue that this kind of spatial modeling is not ideal for audio which is temporally sequential by nature. 
Second, these models are trained on samples with a fixed sampling rate and can only infer at that rate. Inputs with higher or lower rates must be resampled, which inevitably introduces information loss.


In this work, we address these limitations by proposing \textbf{ECHO}, an audio foundation model that uses fr\textbf{E}quen\textbf{C}y-aware \textbf{H}ierarchical enc\textbf{O}ding for variable-length signals.
ECHO is a general-purpose foundation model for machine signals trained on large-scale audio corpus, achieving state-of-the-art performance across various benchmarks, including few-shot anomaly sound detection tasks at DCASE, vibration-based fault detection datasets, as well as multi-modal machine condition monitoring datasets. The main contributions of this work are summarized as follows:
\begin{itemize}
  \vspace{-1mm}\item[i.] frequency-aware band-splitting strategy: splitting spectrogram into frequency sub-bands with a relative frequency positional embedding mechanism tailored for arbitrary sampling rates and frequency resolutions, enabling the model to encode explicit positional context of sub-bands within the full spectrum;
  \vspace{-1mm}\item[ii.] a sliding patch design within each sub-band that suits variable-length signal inputs;
  \vspace{-1mm}\item[iii.] a scalable training framework capable of handling diverse machine signal modalities within a unified representation space; and
  \vspace{-1mm}\item[iv.] achieving state-of-the-art performance on an open-sourced benchmark SIREN for general machine signal embeddings evaluation.
\end{itemize}\vspace{-1mm}


\vspace{-4mm}\section{Related Work}\vspace{-2mm}
\label{sec:related_work}
Concurrent with our study, the FISHER model~\cite{FISHER} introduced band-splitting to handle multi-modal signals with varying sampling rates, where each sub-band is modeled independently using a ViT backbone. Although both works share a similar high-level motivation, they were developed independently and adopt substantially different designs, as reflected in the released code bases\footnote{Our implementation: \url{https://github.com/yucongzh/ECHO}}\footnote{FISHER code: \url{https://github.com/jianganbai/FISHER}}. In contrast to FISHER, our approach incorporates frequency positional encoding within each sub-band before feeding them into the ViT, enabling explicit frequency-aware modeling. Moreover, instead of patch-based image-style tokenization—which is less suitable for variable-length or streaming inputs—our method employs a sliding-window strategy within each sub-band. Conceptually, our framework is designed to simultaneously support variable-length and variable-sampling-rate signals, making it naturally extendable to streaming scenarios.

\vspace{-2.5mm}\section{Model Architecture}\vspace{-2mm}
In this section, we introduce our proposed framework \textbf{ECHO} 
which is designed for robust representation learning of variable-length signals under arbitrary sampling rates. The overall architecture is illustrated in Fig.~\ref{fig:framework}. ECHO consists of four key components: (1) spectrogram extraction, (2) frequency-aware sub-band splitting, (3) temporal sliding patches extraction, and (4) hierarchical encoding. 

\vspace{-3mm}\subsection{Spectrogram Extraction}\vspace{-2mm}
Given an input waveform sampled at frequency $f_s$, we compute its Short-Time Fourier Transform~(STFT) using a predefined window length $t_{win}$ and hop length $t_{hop}$ specified in seconds. We use the magnitude spectrogram. Because these time durations are converted per signal to integer samples, the spectrogram frame rate is fixed by the chosen hop and thus independent of $f_s$. Consequently, for inputs of equal duration, the resulting spectrograms contain the same number of time frames across sampling rates.

\vspace{-3mm}\subsection{Frequency-Aware Sub-band Splitting with Positional Encoding}\vspace{-2mm}
The spectrogram $S$ is uniformly split along the frequency axis into a set of sub-bands with no overlaps, with the number of sub-bands proportional to the sampling rate $f_s$.  
For the $k$-th sub-band spanning $b_{\text{start}}$ to $b_{\text{end}}-1$, the center frequency $f_c$, its normalized position $p$, and the corresponding positional encoding $\text{PE}(p, j)$ are computed as
\begin{equation}
\label{eq:pe}
\begin{aligned}
    f_c &= \frac{(b_{\text{start}} + b_{\text{end}} - 1)}{2} \cdot \frac{f_s}{N_{\text{FFT}}}, \quad 
    p = \frac{f_c}{f_s/2}, \\
    \text{PE}(p, j) &=
    \begin{cases}
        \sin \!\left( \tfrac{\gamma \cdot p}{10000^{2i/d}} \right), & j = 2i, \\
        \cos \!\left( \tfrac{\gamma \cdot p}{10000^{2i/d}} \right), & j = 2i+1 ,
    \end{cases}
\end{aligned}
\end{equation}
where $N_{\text{FFT}}$ is calculated as $f_s\times t_w$, $d$ is the embedding dimension and $\gamma$ is the scaling factor. This design ensures that sub-bands from different sampling rates, but at equivalent relative frequency positions, share consistent positional encodings.

\vspace{-3mm}\subsection{Temporal Sliding Patch Extraction}\vspace{-2mm}
To model signals of variable duration, each sub-band undergoes temporal segmentation. Specifically, we apply a sliding window of length $L$ (equal to the sub-band width) along the time axis, with a stride of $L/2$ to achieve $50\%$ overlap ping.

This operation is efficiently implemented via a two-dimensional convolution with kernel size $(\text{sub-band height}, L)$ and stride $(\text{sub-band height}, L/2)$. The convolution collapses the frequency dimension, resulting in a patch sequence with shape $(N, D)$, where $N$ is the number of temporal patches and $D$ is the channel dimension (i.e., the patch embedding size). Each patch thus represents a localized temporal feature of the sub-band.

\vspace{-3.5mm}\subsection{Hierarchical Encoding}\vspace{-2mm}
Each frequency-aware patch sequence, prepended with a learnable classification~(CLS) token, is fed to the ViT backbone. The CLS token summarizes sub-band information, and the final embedding concatenates all sub-band CLS tokens. This hierarchical design enables ECHO to capture local temporal dependencies within sub-bands while distinguishing frequency ranges via frequency-aware splitting.

\vspace{-3.5mm}\subsection{Training and Inference}\vspace{-2mm}
We adopt a teacher–student framework from EAT~\cite{EAT}.  
During training, each frequency-aware sub-band is treated independently. The student receives masked inputs, while the teacher is updated via Exponential Moving Average~(EMA) of the student:
$
    \theta_{\text{teacher},t} = \alpha \theta_{\text{teacher},t-1} + (1-\alpha)\theta_{\text{student},t},
$
with momentum $\alpha$. We employ two self-supervised objectives: (1) global alignment between the temporal mean of the teacher’s layer outputs and student CLS token, and (2) frame-level alignment on masked positions. This dual-level supervision enforces consistency at both coarse and fine scales.

During inference, the full spectrogram is processed by ECHO. CLS tokens from all $K$ sub-bands are concatenated into a hierarchical embedding
$
    \mathbf{z} = [\text{CLS}_1, \dots, \text{CLS}_K]
$
for downstream tasks.

\begin{table}[t]
\centering
\vspace{-2.5mm}\caption{SIREN Benchmark datasets and their characteristics. SR: Sample rate; MAFAULDA: Machinery Fault Database; CWRU: CWRU Bearing dataset; IIEE: IDMT Electric Engine dataset; IICA: IDMT Compressed Air dataset.}\vspace{0.5mm}
\label{tab:siren_datasets}
\resizebox{\columnwidth}{!}{\Large
\begin{tabular}{c c c c c c}
\toprule
Dataset & Modality & SR & \#Classes & Split & Scoring \\
\midrule
DCASE2020~\cite{dcase2020} & Sound & 16k & 2 & official & KNN \\
DCASE2021~\cite{dcase2021} & Sound & 16k & 2 & official & KNN \\
DCASE2022~\cite{dcase2022} & Sound & 16k & 2 & official & KNN \\
DCASE2023~\cite{dcase2023} & Sound & 16k & 2 & official & KNN \\
DCASE2024~\cite{dcase2024} & Sound & 16k & 2 & official & KNN \\
DCASE2025~\cite{dcase2025} & Sound & 16k & 2 & official & KNN \\
MAFAULDA~\cite{mafaulda}  & Sound/Vibration & 50k & 10 & LOOCV & KNN \\
CWRU~\cite{cwru}      & Vibration & 12k & 10 & LOOCV & KNN \\
IIEE~\cite{iiee}      & Sound & 44.1k & 3 & official & KNN \\
IICA~\cite{iica}      & Sound & 48k & 3 & 5-Fold-CV & KNN \\
\bottomrule
\end{tabular}
}
\vspace{-6mm}
\end{table}

\begin{table*}[t]
    \centering
    \caption{Performance~(\%) comparison of pre-trained foundation models across DCASE challenges and machine fault diagnosis datasets in our SIREN benchmark~(k=5). DCASE tasks are reported by the (harmonic) means of AUCs and partial AUCs among machines, while fault classification tasks are reported by accuracy. DCASE tasks are evaluated according to the officials. ``Mean'' stands for arithmetic mean. Sample rates are listed under dataset names. FS$^*$: Freesound derived from WavCaps~\cite{mei2023wavcaps}.}\vspace{-2mm}
    \resizebox{\textwidth}{!}{\Large
    \begin{threeparttable}
    \label{tab:performance_comparison}
    \begin{tabular}{l ccc|cccccc c|cccc c|c}
    \toprule
    \multirow{3}{*}{Model} & \multirow{3}{*}{Datasets} & \multirow{3}{*}{Scale} & \multirow{3}{*}{\#Param.} & \multicolumn{7}{c|}{DCASE Tasks} & \multicolumn{5}{c|}{Fault Classification Tasks} & \multirow{3}{*}{Mean} \\
    \cline{5-11} \cline{12-16}
     & & & & 2020 & 2021 & 2022 & 2023 & 2024 & 2025 & Mean & IIEE & IICA & CWRU & MAFAULDA & Mean & \\
    \cline{5-11} \cline{12-16}
     & & & & 16k & 16k & 16k & 16k & 16k & 16k & - & 44.1k & 48k & 12k & 50k & - & \\
    \midrule
    BEATs~\cite{BEATs}    & \multirow{1}{*}{AS2M} 
       & Base & 90M & \textbf{74.26} & \textbf{61.31} & 58.97 & 62.89 & 55.89 & 57.84 & \underline{61.86} & 65.81 & 91.55 & 88.57 & 99.69/63.66 & 81.86 & 71.86 \\
    \midrule
    \multirow{4}{*}{CED~\cite{CED}} & \multirow{4}{*}{AS2M} 
       & Base  & 86M  & 67.75 & 56.67 & 57.26 & 60.84 & 57.83 & 57.72 & 59.68 & 80.21 & 86.08 & 81.90 & 99.74/66.48 & 82.88 & 71.28 \\
     & & Small & 22M  & 67.69 & 56.66 & 56.79 & 60.04 & 57.24 & 57.89 & 59.39 & 74.42 & 85.66 & 85.71 & 99.59/64.43 & 81.96 & 70.67 \\
     & & Mini  & 10M & 67.59 & 56.35 & 57.03 & 59.85 & 56.39 & 57.43 & 59.11 & 74.17 & 84.91 & 82.86 & 99.64/63.66 & 81.05 & 70.08 \\
     & & Tiny  & 5.5M & 67.21 & 56.17 & 56.77 & 59.61 & 56.19 & 57.82 & 58.96 & 72.74 & 84.02 & 82.86 & 99.64/63.51 & 80.55 & 69.76 \\
    \midrule
    \multirow{2}{*}{EAT~\cite{EAT}} & \multirow{2}{*}{AS2M} 
       & Large & 0.3B & \underline{73.94} & 57.47 & 58.54 & 61.66 & \textbf{57.89} & \textbf{60.01} & 61.58 & 68.33 & 89.96 & \textbf{91.43} & 99.33/85.03 & 87.72 & 74.28 \\
     & & Base  & 86M  & 72.13 & 57.79 & 58.57 & 59.69 & 57.12 & \underline{59.75} & 60.84 & 78.97 & 89.01 & 85.71 & 99.90/84.52 & 86.71 & 74.15 \\
    \midrule
    \multirow{3}{*}{Dasheng~\cite{Dasheng}} & \multirow{3}{*}{\begin{tabular}{@{}c@{}}AS2M+MTG\\+VGG\\+ACAV\end{tabular}} 
       & 1.2B & 1.2B & 69.48 & 57.06 & 57.29 & 61.23 & 57.13 & 57.12 & 59.88 & 96.09 & 91.91 & 90.48 & 99.69/77.24 & 91.08 & 75.48 \\
     & & 0.6B & 0.6B & 68.18 & 56.76 & 56.61 & 59.94 & 56.75 & 56.73 & 59.16 & 99.11 & 92.11 & 89.52 & 99.74/78.22 & 91.74 & 75.45 \\
     & & Base & 86M  & 69.15 & 57.27 & 57.87 & 60.70 & 57.71 & 57.01 & 59.95 & 99.36 & 90.88 & 88.57 & 99.85/81.96 & 92.12 & 76.04 \\
    \midrule
    \multirow{3}{*}{FISHER~\cite{FISHER}} & \multirow{3}{*}{\begin{tabular}{@{}c@{}}AS2M+MTG\\+M4A+FS\end{tabular}} 
       & Small & 22M  & 70.54 & 59.51 & \underline{59.79} & 61.83 & 55.66 & 58.68 & 61.00 & 97.48 & 94.20 & 86.67 & 100.0/85.29 & \underline{92.73} & 76.86 \\
     & & Mini  & 10M & 69.98 & 58.39 & 57.91 & 60.35 & 55.91 & 56.90 & 59.91 & \underline{99.90} & \underline{94.50} & 73.33 & \underline{100.0/87.75} & 91.10 & 75.50 \\
     & & Tiny  & 5.5M & 70.64 & 58.51 & 57.11 & 58.46 & 55.34 & 57.69 & 59.62 & 99.80 & \textbf{95.43} & 75.24 & \textbf{100.0/88.52} & 91.80 & 75.71 \\
    \midrule
    \multirow{2}{*}{ECHO} & \multirow{2}{*}{\begin{tabular}{@{}c@{}}AS2M+MTG\\+FS$^*$\end{tabular}} 
       & Small & 22M  & 72.23 & \underline{60.20} & \textbf{59.96} & \underline{63.71} & \underline{57.86} & 58.70 & \textbf{62.11} & 99.85 & 93.67 & \underline{90.48} & 99.54/82.42 & \textbf{93.19} & \textbf{77.65} \\    
     & & Tiny  & 5.5M & 70.14 & 59.01 & 59.76 & \textbf{63.75} & 56.91 & 58.40 & 61.33 & \textbf{100.0} & 93.58 & 90.48 & 99.85/78.83 & 92.55 & \underline{76.94} \\       
    \bottomrule
    \end{tabular}\vspace{-1mm}
    \end{threeparttable}
    }\vspace{-5mm}
\end{table*}

\vspace{-3mm}\section{Benchmark for Evaluation}\vspace{-3mm}
For fair comparison, we open-source an evaluation benchmark called SIREN~(\textbf{SI}gnal \textbf{R}epresentation \textbf{E}valuatio\textbf{N} toolkit). SIREN is tailored for general signal diagnosis, including tasks like few-shot anomalous detection (DCASE Task 2 series), and machine fault diagnosis/classification. The detailed information of the datasets and the corresponding tasks are shown in Table~\ref{tab:siren_datasets} and the GitHub repository\footnote{Codes available at \url{https://github.com/yucongzh/SIREN}}. The evaluation protocol and metrics are as follows.

\vspace{1mm}\noindent\textbf{DCASE Task 2 series:} We follow the official development/evaluation splits~\cite{dcase2020,dcase2021,dcase2022,dcase2023,dcase2024,dcase2025}, computing file-level anomaly scores, and aggregate them with machine/section/domain grouping. Per year, we report ROC-AUC and partial AUC (pAUC), and summarize performance by harmonic means of AUC and pAUC across development and evaluation sets.  

\vspace{1mm}\noindent\textbf{Fault classification:} For datasets~(MAFAULDA~\cite{mafaulda}, CWRU~\cite{cwru}, IICA~\cite{iica}) without official train-test split, we use cross validation~(CV) for evaluation. On MAFAULDA and CWRU, we use leave-one-out CV~(LOOCV) due to limited data for each fault class; on IICA, we use 5-fold CV. For IIEE~\cite{iiee}, we use the given train-test split for evaluation. These settings (see Table~\ref{tab:siren_datasets}) balance reliability and computational cost. Accuracy is reported for each dataset. 

\vspace{1mm}\noindent\textbf{Scoring.}
We use $k$-nearest neighbors (k-NN): (1) compute training embeddings $\{e_i\}_{i=1}^{N}$ to build a memory bank $\mathcal{M}$; (2) for a test embedding $e^\ast$, retrieve its $k$ nearest neighbors in $\mathcal{M}$. For DCASE anomaly detection, the score is the nearest-neighbor distance ($k{=}1$). For fault classification, the label is the majority vote of the $k$ neighbors.

\vspace{-3mm}\section{Experiment}
\label{sec:exp}

\vspace{-2mm}\subsection{Implementation Details}\vspace{-2mm}
We adopt the same backbone architecture as the ViT. Currently, we have released two editions of ECHO: small and tiny. The spectrogram is extracted using a window size of 25 ms and window shift of 10 ms on normalized raw signals. The sub-band width in our model is fixed to 32. 
The model is trained for a total of 400,000 steps using four NVIDIA GeForce RTX 3090 GPUs with a global batch size of 256, with 4,000 warm-up steps. The learning rate is scaled relative to the effective batch size~\cite{audioMAE}, with a base learning rate of $10^{-4}$. 
Training employs a cosine learning rate scheduler with a linear warm-up phase spanning 40,000 steps. The minimum learning rate is set to $10^{-5}$, and weight decay is applied with a coefficient of 0.05. 

\vspace{-3.5mm}\subsection{Baseline Models}\vspace{-2mm}
Currently, we include 5 pre-trained foundation models for comparison: BEATs~\cite{BEATs}, CED~\cite{CED}, EAT~\cite{EAT}, Dasheng~\cite{Dasheng}, and FISHER~\cite{FISHER}. All models use ViT-style structure as the backbone model, and are trained using open-sourced audio datasets across various domains, including full AudioSet~(AS2M)~\cite{gemmeke2017audio}, MTG-Jamendo~(MTG)~\cite{bogdanov2019mtg}, VGGSound~(VGG)~\cite{chen2020vggsound}, Music4all~(M4A)~\cite{santana2020music4all}, ACAV~\cite{lee2021acav100m}, and Freesound~(FS)\footnote{\url{https://freesound.org/}}. 

\vspace{-3mm}\subsection{Experimental Results}\vspace{-2mm}
Table~\ref{tab:performance_comparison} reports the performance on the SIREN benchmark. Several observations can be made:

\vspace{0.5mm}\noindent\textbf{1) Effect of dataset scaling.} By comparing BEATs (71.86\%), CED (71.28\%) and EAT (74.15\%) with Dasheng (76.04\%) with base scale, we observe that incorporating additional large-scale datasets might help mitigate the domain mismatch between general audio pre-training and machine sound analysis. 
A similar trend can also be found in FISHER (76.86\%) and our proposed ECHO (77.65\%), both trained on additional audio datasets, showing that scale-up of training data consistently enhances cross-domain representation learning.

\vspace{0.5mm}\noindent\textbf{2) Sliding-patch vs. conventional patch modeling.} Traditional foundation models (BEATs, CED, and EAT) rely on conventional patch tokenization, yielding total average scores in the range of 70–74\% in our SIREN benchmark. In contrast, Dasheng introduces a sliding-patch strategy, leading to a higher overall mean of 76.04\%, outperforming EAT (74.15\%) by $+1.89\%$. A similar observation is confirmed within the band-splitting family: our proposed ECHO reaches a total average of 77.65\% on SIREN, surpassing FISHER (76.86\%) by $+0.79\%$. These results highlight that sliding-patch modeling is more effective than fixed patch partitioning for machine sound analysis.

\vspace{0.5mm}\noindent\textbf{3) Band-splitting architecture.} Compared with conventional pre-trained foundation models, band-splitting-based methods achieve superior performance in fault classification across multiple modalities and sampling conditions, while maintaining competitive results on DCASE tasks. As shown in Table~\ref{tab:performance_comparison}, ECHO (93.19\%) and FISHER (92.73\%) rank first and second on fault classification tasks, with model with small scale. This confirms that decomposing signals into frequency sub-bands facilitates robust cross-sampling-rate anomaly detection. Furthermore, ECHO consistently outperforms FISHER, which we attribute to the integration of frequency positional encoding, enabling better frequency-aware modeling across sub-bands. 
It is noteworthy that Dasheng (92.12\%), trained only on samples with fixed sampling rate, also exhibits strong adaptability to diverse fault classification tasks, a capability that may stem from its use of large-scale training data.

\vspace{0.5mm}\noindent\textbf{4) Effect of model scaling.} Comparing ECHO-Small (77.65\%) and ECHO-Tiny (76.94\%) shows that enlarging the model scale yields consistent improvements across both DCASE (62.11\% vs. 61.33\%) and fault classification tasks (93.19\% vs. 92.55\%). This indicates that our architecture retains scalability potential, and larger variants may further boost generalization. 

Overall, ECHO achieves the highest overall performance (77.65\%) in our SIREN benchmark, outperforming the strong baseline (FISHER, 76.86\%) by $+0.79\%$. These results show that combining frequency-aware band-splitting with sliding-patch modeling is effective in our principled framework for cross-domain machine signal representation learning.


\vspace{-3.5mm}
\section{Conclusion}
\vspace{-3mm}
In this article, we proposed ECHO, a novel pre-trained foundation model that incorporates a frequency-aware band-splitting strategy to handle diverse sampling rates and uses sliding patches to accommodate inputs of varying lengths. The model was evaluated on the newly introduced open-source benchmark SIREN, which includes tasks such as anomaly detection and fine-grained anomaly classification across multiple machine modalities, including acoustics and vibrations. Experimental results demonstrate that ECHO achieves state-of-the-art performance in both anomaly detection and fault classification, highlighting its potential for a wide range of industrial anomaly detection applications.


{
\small
\bibliographystyle{IEEEbib}
\bibliography{refs}

\begin{thebibliography}{10}
\providecommand{\url}[1]{#1}
\csname url@samestyle\endcsname
\providecommand{\newblock}{\relax}
\providecommand{\bibinfo}[2]{#2}
\providecommand{\BIBentrySTDinterwordspacing}{\spaceskip=0pt\relax}
\providecommand{\BIBentryALTinterwordstretchfactor}{4}
\providecommand{\BIBentryALTinterwordspacing}{\spaceskip=\fontdimen2\font plus
\BIBentryALTinterwordstretchfactor\fontdimen3\font minus \fontdimen4\font\relax}
\providecommand{\BIBforeignlanguage}[2]{{%
\expandafter\ifx\csname l@#1\endcsname\relax
\typeout{** WARNING: IEEEtran.bst: No hyphenation pattern has been}%
\typeout{** loaded for the language `#1'. Using the pattern for}%
\typeout{** the default language instead.}%
\else
\language=\csname l@#1\endcsname
\fi
#2}}
\providecommand{\BIBdecl}{\relax}
\BIBdecl

\bibitem{yan2024comprehensive}
P.~Yan, A.~Abdulkadir, P.-P. Luley, M.~Rosenthal, G.~A. Schatte, B.~F. Grewe, and T.~Stadelmann, ``A comprehensive survey of deep transfer learning for anomaly detection in industrial time series: Methods, applications, and directions,'' \emph{IEEE Access}, vol.~12, pp. 3768--3789, 2024.

\bibitem{panns}
Q.~Kong, Y.~Cao, T.~Iqbal, Y.~Wang, W.~Wang, and M.~D. Plumbley, ``Panns: Large-scale pretrained audio neural networks for audio pattern recognition,'' \emph{IEEE/ACM Trans. ASLP}, vol.~28, pp. 2880--2894, 2020.

\bibitem{audioMAE}
P.-Y. Huang, H.~Xu, J.~Li, A.~Baevski, M.~Auli, W.~Galuba \emph{et~al.}, ``Masked autoencoders that listen,'' in \emph{Proc. NeurIPS}, vol.~35, 2022, pp. 28\,708--28\,720.

\bibitem{koutini22_interspeech}
K.~Koutini, J.~Schlüter, H.~Eghbal-zadeh, and G.~Widmer, ``Efficient training of audio transformers with patchout,'' in \emph{{Proc. Interspeech}}, {2022}, pp. {2753--2757}.

\bibitem{BEATs}
S.~Chen, Y.~Wu, C.~Wang, S.~Liu, D.~Tompkins, Z.~Chen \emph{et~al.}, ``{BEAT}s: Audio pre-training with acoustic tokenizers,'' in \emph{Proc. ICML}, 2023, pp. 5178--5193.

\bibitem{CED}
H.~Dinkel, Y.~Wang, Z.~Yan, J.~Zhang, and Y.~Wang, ``{CED}: Consistent ensemble distillation for audio tagging,'' in \emph{Proc. ICASSP}, 2024, pp. 291--295.

\bibitem{EAT}
W.~Chen, Y.~Liang, Z.~Ma, Z.~Zheng, and X.~Chen, ``{EAT}: Self-supervised pre-training with efficient audio transformer,'' in \emph{Proc. IJCAI}, 2024, pp. 3807--3815.

\bibitem{ViT}
A.~Dosovitskiy, L.~Beyer, A.~Kolesnikov, D.~Weissenborn, X.~Zhai, T.~Unterthiner \emph{et~al.}, ``An image is worth 16x16 words: Transformers for image recognition at scale,'' in \emph{Proc. ICLR}, 2021.

\bibitem{Dasheng}
H.~Dinkel, Z.~Yan, Y.~Wang, J.~Zhang, Y.~Wang, and B.~Wang, ``Scaling up masked audio encoder learning for general audio classification,'' in \emph{Proc. Interspeech}, 2024, pp. 547--551.

\bibitem{jiang24c_interspeech}
A.~Jiang, B.~Han, Z.~Lv, Y.~Deng, W.-Q. Zhang, X.~Chen \emph{et~al.}, ``Anopatch: Towards better consistency in machine anomalous sound detection,'' in \emph{Proc. Interspeech}, 2024, pp. 107--111.

\bibitem{zheng2024improving}
X.~Zheng, A.~Jiang, B.~Han, Y.~Qian, P.~Fan, J.~Liu, and W.-Q. Zhang, ``Improving anomalous sound detection via low-rank adaptation fine-tuning of pre-trained audio models,'' in \emph{Proc. SLT}, 2024, pp. 969--974.

\bibitem{han2025exploring}
B.~Han, A.~Jiang, X.~Zheng, W.-Q. Zhang, J.~Liu, P.~Fan, and Y.~Qian, ``Exploring self-supervised audio models for generalized anomalous sound detection,'' \emph{IEEE Trans. ASLP}, 2025.

\bibitem{FISHER}
P.~Fan, A.~Jiang, S.~Zhang, Z.~Lv, B.~Han, X.~Zheng \emph{et~al.}, ``{FISHER}: A foundation model for multi-modal industrial signal comprehensive representation,'' \emph{arXiv preprint arXiv:2507.16696}, 2025.

\bibitem{dcase2020}
Y.~Koizumi, Y.~Kawaguchi, K.~Imoto, T.~Nakamura, Y.~Nikaido, R.~Tanabe \emph{et~al.}, ``Description and discussion on dcase2020 challenge task2: Unsupervised anomalous sound detection for machine condition monitoring,'' in \emph{Proc. DCASE}, 2020, pp. 81--85.

\bibitem{dcase2021}
Y.~Kawaguchi, K.~Imoto, Y.~Koizumi, N.~Harada, D.~Niizumi, K.~Dohi \emph{et~al.}, ``Description and discussion on dcase 2021 challenge task 2: Unsupervised anomalous detection for machine condition monitoring under domain shifted conditions,'' in \emph{Proc. DCASE}, 2021, pp. 186--190.

\bibitem{dcase2022}
K.~Dohi, K.~Imoto, N.~Harada, D.~Niizumi, Y.~Koizumi, T.~Nishida \emph{et~al.}, ``Description and discussion on dcase 2022 challenge task 2: Unsupervised anomalous sound detection for machine condition monitoring applying domain generalization techniques,'' in \emph{Proc. DCASE}, 2022.

\bibitem{dcase2023}
------, ``Description and discussion on {DCASE} 2023 challenge task 2: First-shot unsupervised anomalous sound detection for machine condition monitoring,'' in \emph{Proc. DCASE}, 2023, pp. 31--35.

\bibitem{dcase2024}
T.~Nishida, N.~Harada, D.~Niizumi, D.~Albertini, R.~Sannino, S.~Pradolini \emph{et~al.}, ``Description and discussion on dcase 2024 challenge task 2: First-shot unsupervised anomalous sound detection for machine condition monitoring,'' in \emph{Proc. DCASE}, 2024, pp. 111--115.

\bibitem{dcase2025}
------, ``Description and discussion on dcase 2025 challenge task 2: First-shot unsupervised anomalous sound detection for machine condition monitoring,'' \emph{arXiv preprint arXiv:2506.10097}, 2025.

\bibitem{mafaulda}
{Signals, Multimedia and Telecommunications Laboratory (SMT)}, ``{MAFAULDA}: Machinery fault database,'' SMT Lab. [Online]. Available: \url{http://www02.smt.ufrj.br/~offshore/mfs/page_01.html}.

\bibitem{cwru}
{Case Western Reserve University}, ``Bearing data center: Seeded fault test data,'' Case School of Engineering. [Online]. Available: \url{https://engineering.case.edu/bearingdatacenter}.

\bibitem{iiee}
S.~Grollmisch, J.~Abe{\ss}er, J.~Liebetrau, and H.~Lukashevich, ``Sounding industry: Challenges and datasets for industrial sound analysis,'' in \emph{Proc. EUSIPCO}, 2019, pp. 1--5.

\bibitem{iica}
D.~Johnson, J.~Kirner, S.~Grollmisch, and J.~Liebetrau, ``Compressed air leakage detection using acoustic emissions with neural networks,'' in \emph{Proc. Inter-Noise}, Seoul, South Korea, 2020, pp. 5662--5673.

\bibitem{mei2023wavcaps}
X.~Mei, C.~Meng, H.~Liu, Q.~Kong, T.~Ko, C.~Zhao \emph{et~al.}, ``Wavcaps: A chatgpt-assisted weakly-labelled audio captioning dataset for audio-language multimodal research,'' \emph{IEEE/ACM Trans. ASLP}, vol.~32, pp. 3339--3354, 2024.

\bibitem{gemmeke2017audio}
J.~F. Gemmeke, D.~P. Ellis, D.~Freedman, A.~Jansen, W.~Lawrence, R.~C. Moore \emph{et~al.}, ``Audio set: An ontology and human-labeled dataset for audio events,'' in \emph{Proc. ICASSP}, 2017, pp. 776--780.

\bibitem{bogdanov2019mtg}
D.~Bogdanov, M.~Won, P.~Tovstogan, A.~Porter, and X.~Serra, ``The mtg-jamendo dataset for automatic music tagging,'' in \emph{Proc. ICML}, 2019.

\bibitem{chen2020vggsound}
H.~Chen, W.~Xie, A.~Vedaldi, and A.~Zisserman, ``Vggsound: A large-scale audio-visual dataset,'' in \emph{Proc. ICASSP}, 2020, pp. 721--725.

\bibitem{santana2020music4all}
I.~A.~P. Santana, F.~Pinhelli, J.~Donini, L.~Catharin, R.~B. Mangolin, V.~D. Feltrim \emph{et~al.}, ``Music4all: A new music database and its applications,'' in \emph{Proc. IWSSIP}, 2020, pp. 399--404.

\bibitem{lee2021acav100m}
S.~Lee, J.~Chung, Y.~Yu, G.~Kim, T.~Breuel, G.~Chechik, and Y.~Song, ``Acav100m: Automatic curation of large-scale datasets for audio-visual video representation learning,'' in \emph{Proc. ICCV}, 2021, pp. 10\,274--10\,284.

\end{thebibliography}
}

\end{document}